\documentclass[conference, lettersize]{IEEEtran}
\IEEEoverridecommandlockouts
\normalsize
\usepackage{cite}
\usepackage[normalem]{ulem}
\ifCLASSINFOpdf
   \usepackage[pdftex]{graphicx}
   \DeclareGraphicsExtensions{.pdf}
\else
   \usepackage[dvips]{graphicx}
   \DeclareGraphicsExtensions{.eps}
\fi

 \ifCLASSOPTIONcompsoc
	\usepackage[caption=false,font=footnotesize,labelfont=sf,textfont=sf]{subfig}
\else
	\usepackage[caption=false,font=footnotesize]{subfig}
\fi

\RequirePackage{lineno}

\usepackage{color}
\usepackage[usenames]{xcolor}
\usepackage{microtype}

\usepackage{amsmath,amssymb,amsfonts,bbm}
\usepackage{booktabs}
\usepackage[inline]{enumitem}
\usepackage{graphicx, float}
\usepackage{nccmath}
\usepackage{multirow}

\usepackage[backgroundcolor=cyan,bordercolor=gray,textsize=small,textwidth=5.8cm]{todonotes}

\usepackage{stfloats}

\usepackage{url}

\usepackage[ruled,vlined,linesnumbered]{algorithm2e}
\usepackage{algpseudocode}
\usepackage[flushleft]{threeparttable}

\makeatletter
\def\BState{\State\hskip-\ALG@thistlm}
\makeatother

\usepackage[backgroundcolor=cyan,bordercolor=gray,textsize=small,textwidth=5.8cm]{todonotes}

\makeatletter
\def\BState{\State\hskip-\ALG@thistlm}
\makeatother

\usepackage{textcomp}

\def\BibTeX{{\rm B\kern-.05em{\sc i\kern-.025em b}\kern-.08em T\kern-.1667em\lower.7ex\hbox{E}\kern-.125emX}}

\newcommand{\vertiii}[1]{{\left\vert\kern-0.25ex\left\vert\kern-0.25ex\left\vert #1
    \right\vert\kern-0.25ex\right\vert\kern-0.25ex\right\vert}}

\newlist{myitemize}{itemize}{3}
\setlist[myitemize]{noitemsep, topsep=0pt}
\setlist[myitemize,1]{label=\protect\itemcircle ,leftmargin=1.0in}
\setlist[myitemize,2]{label=$\rightarrow$,leftmargin=1em}
\setlist[myitemize,3]{label=$\diamond$}

\usepackage{tcolorbox}
\definecolor{BlueGray}{HTML}{546E7A}
\definecolor{BlueGrayLight}{HTML}{607D8B}
\definecolor{Amber}{HTML}{FF8F00}
\definecolor{Orange}{HTML}{FB8C00}
\definecolor{DeepOrange}{HTML}{FF7043}
\definecolor{Indigo}{HTML}{3949AB}
\definecolor{Teal}{HTML}{00897B}
\definecolor{LightGreen}{HTML}{7CB342}
\definecolor{LightBlue}{HTML}{0277BD}
\definecolor{Purple}{HTML}{AB47BC}
\definecolor{DeepPurple}{HTML}{7E57C2}

\begin{document}

\newcommand{\roa}[1]{\textcolor{blue}{[Ramoni: #1]}}

\title{
Multi-User Beamforming with Deep Reinforcement Learning in Sensing-Aided Communication\\
\thanks{The work by Xiyu Wang, Gilberto Berardinelli, Petar Popovski and Ramoni Adeogun are supported by the HORIZON JU-SNS-2022-STREAM-B-01-02 CENTRIC project (Grant Agreement No.101096379). Xiyu Wang is also supported in part by the European Union’s Horizon 2020 research and innovation programme under the Marie Skłodowska-Curie Grant agreement No. 101146247. Hei Victor Cheng's work is supported by the Aarhus Universitets Forskningsfond project No. AUFF39001.}
}\author{\IEEEauthorblockN{Xiyu Wang$^1$, Gilberto Berardinelli$^1$, Hei Victor Cheng$^2$, Petar Popovski$^1$, Ramoni Adeogun$^1$}
\IEEEauthorblockA{$^1$ Department of Electronic Systems, Aalborg University, Denmark \\
$^2$ Department of Electrical and Computer Engineering, Aarhus University, Denmark}
Email: xiyuw@es.aau.dk, gb@es.aau.dk, hvc@ece.au.dk, petarp@es.aau.dk, ra@es.aau.dk
}

\maketitle

\begin{abstract}
   Mobile users are prone to experience beam failure due to beam drifting in millimeter wave (mmWave) communications. Sensing can help alleviate beam drifting with timely beam changes and low overhead since it does not need user feedback. This work studies the problem of 
   optimizing sensing-aided communication by dynamically managing beams allocated to mobile users. A multi-beam scheme is introduced, which  allocates multiple beams to the users that need an update on the angle of departure (AoD) estimates and a single beam to the users that have satisfied AoD estimation precision. A deep reinforcement learning (DRL) assisted method is developed to optimize the beam allocation policy, relying only upon the sensing echoes. For comparison, a heuristic AoD-based method using approximated Cramér-Rao lower bound (CRLB) for allocation is also presented. Both methods require neither user feedback nor prior state evolution information.  Results show that the DRL-assisted method achieves a considerable gain in throughput than the conventional beam sweeping method and the AoD-based method, and it is robust to different user speeds.     
\end{abstract}

\section{Introduction}
Sensing gains increasing importance in sixth-generation (6G) communication as it provides more accurate and higher amounts of information for assisting communication. As such, integrated communication and sensing (ISAC) become one crucial component of the 6G technologies for improving the quality of services and expanding application scenarios.  Nonetheless, sensing shares resources with communication, and how to manage resource allocation between the two modules attracts interest in ISAC \cite{liu_integrated_2022}. 

Allocating resources in the beam space is promising due to the maturation of the millimeter wave (mmWave) space division multiplexing \cite{xue_survey_2024}. Moreover, it enhances spectrum efficiency by offering a new dimension for resource allocation with respect to time and frequency \cite{zhuo_2024_multibeam_isac}. 
However, the narrow beams in the mmWave spectrum suffer from beam drifting, that is, the beamforming target is not aligned with the center of the beam, especially in a high-mobility environment \cite{Zhang2021BeamDrift}. In a dynamic environment, beams need to be reallocated timely to prevent beam failure. This work focuses on multi-beam management that alleviates beam drifting and improves communication performance in a dynamic environment. 

Various frameworks have been proposed for aligning beams and resolving beam drifting. 3GPP standards adopt periodically sweeping beams as the solution \cite{xue_survey_2024}. The sweeping frequency needs to be optimized according to the system kinematics. The work \cite{Zhang2021BeamDrift} proposes a varying-beamwidth transmission to mitigate the beam drifting for a single-user case. This protocol relies on user feedback which can yield high overhead and result in late beam updates.
Sensing helps understand the wireless environment and thus, can potentially prevent beam drifting more robustly. 
Available works \cite{liu_integrated_2022, MugenPeng2023, chen_multiuser_2023, liu_radar-assisted_2020} deploy a sensing receiver to receive reflected echoes, from which the angle of departure (AoD) is estimated and predicted to improve the beam alignment using state evolution models.
The prior knowledge of state evolution is however difficult to obtain especially in multi-user dynamic environments.

Learning-based algorithms stand out because they can well-capture complex dynamic environments to optimize beam management. Beam selection has been done using deep neural networks to maximize the sum rate \cite{Chen2024}. A deep reinforcement learning (DRL) approach to optimize the beam and power allocation in an ISAC system is studied in \cite{deep_rl_liu_2024_wcnc}. It maximizes the sum rate while taking estimation precision, i.e., Cramér-Rao lower bound (CRLB), as a penalty. However, the resource allocation is optimized for one time instant and the proposed method requires genie information for the CRLB calculation. Management of beamwidth and interference between a communication train and a sensing target in a fast-moving railway scenario is studied in \cite{Li_drl_bm_2022}. The proposed DRL model requires user signal-to-interference-plus-noise ratio (SINR) feedback as observations, which increases the scheduling overhead.

This paper studies the multi-beam management to alleviate beam-drifting for multiple users in a dynamic sensing-assisted communication system. The proposed methods resolve beam drifting without neither prior information nor user feedback, but only relying on the sensing information. In addition, available research on beam management in ISAC primarily focuses on maximizing physical layer communication metrics, such as maximizing sum rate or SINR, to optimize waveform design\cite{liu_radar-assisted_2020, deep_rl_liu_2024_wcnc, Li_drl_bm_2022, chen_multiuser_2023, Chen2024}. This work instead shifts the focus to medium access control (MAC)-layer metric, i.e., packet throughput by jointly optimizing downlink scheduling, power allocation, and beam allocation. 
Specifically, we propose to group users into communication and sensing users dynamically, and use multiple beams for sensing users to broaden the sensing field of view (FoV) and a single beam for communication users. 

The main contributions are: A heuristic AoD-based method is developed that uses the approximated CRLB of AoD to group users and assign beams.
Then a DRL-assisted approach for user scheduling and beam allocation is proposed. The DRL model takes the beamforming outputs of the reflected echo as states and optimizes the policy that maximizes packet throughput. 
Comparative results in mobile scenarios show that the DRL-assisted beam management outperforms the periodical beam sweeping and the heuristic AoD-based method, and is robust to user movements and speed. The proposed methods adjust beams timely and mitigate beam drifting. Furthermore, our approach verifies that sensing helps beam management without prior information or user feedback.


\section{System Model}
\begin{figure}
    \centering
    \begin{tabular}{cc}
        \subfloat[]
    {\includegraphics[width=0.24\textwidth]{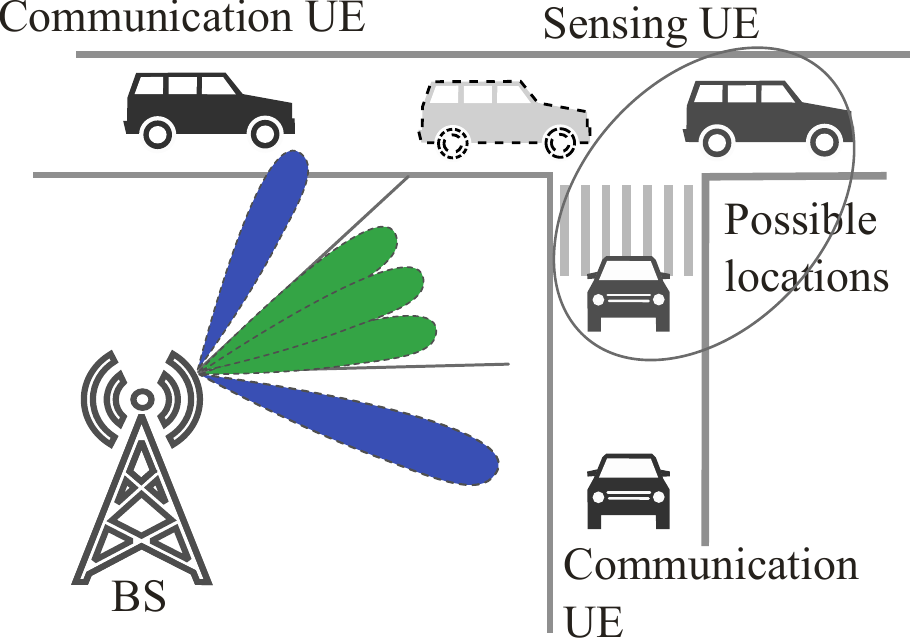}
    \label{fig:scenario}}
        \subfloat[]
    {\includegraphics[width=0.15\textwidth]{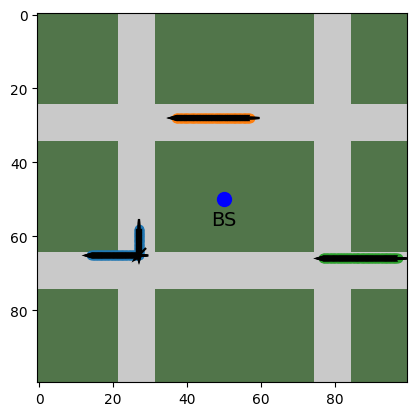}
    \label{fig:street}}
\end{tabular}
    \caption{(a) An illustration of considered multi-user ISAC scenario. (b) A snapshot of user trajectory in one frame. }
    \label{fig:isac_system}\vspace{-10pt}
\end{figure}

Let us consider a multi-user mmWave vehicular communication system in which a base station (BS) equipped with $N_t$ antennas is serving multiple moving users, as shown in Fig.~\ref{fig:scenario}. The BS is also equipped with a sensing receiver with $N_r$ antennas which assists the communication. Both transmitter and receiver are uniform linear arrays (ULA) with $\Delta_r = v_c/2f_c$ separation between antenna elements, where $v_c$ denotes the speed of light and $f_c$ is the carrier frequency. Without loss of generality, we assume $N_r = N_t$, and the number of user equipment (UE) to be much less than the number of BS antennas, i.e., $U \ll N_t$. 

We consider a narrowband wireless system since this work mainly focuses on the angular domain. In order to yield more sensing information, a wideband system is needed. To have a clear elaboration on the method, the extension to wideband system is left for future work.

Consider a frame consisting of $N$ transmission time intervals (TTI). At the beginning of a frame, the initial beam alignment is done by beam sweeping. Specifically, the best beam pairs are formed by BS transmitting pilots at all possible beams and users measuring and reporting to BS\cite{xue_survey_2024}.
We assume that the BS knows which AoD is associated with which user such that the targeted signal can be transmitted in a specific direction. 

\subsection{Transmit Signal}
We consider that the BS selects beams from the effective discrete Fourier transformation (DFT)-based codebook $\tilde{\mathbf{F}}$. The $i$-th codeword in the DFT codebook $\tilde{\mathbf{F}}$ be represented on a uniform grid of $N_t$ points, given by
\begin{equation}
\small
    [\tilde{\mathbf{F}}]_{:,i} = \frac{1}{\sqrt{N_t}}[1, e^{-j \varphi_i}, \cdots, e^{-j(N_t-1)\varphi_i}] .
\label{eq:codebook}
\end{equation}
where $\varphi_i = \pi\frac{2i-1-N_t}{N_t}$. 
During the $n$-th TTI,  the beamforming codeword selected for user $u$ is denoted as $\boldsymbol{f}_{u,n}$, and it stays the same.
The signal transmitted from the BS to all the users at time $t$ during the $n$-th TTI is given by
\begin{equation}
\small
    \boldsymbol{s}_n(t) = \sum_{u=1}^{U} \sqrt{P_{u,n}}\boldsymbol{f}_{u,n} \dot{s}_{u,n}(t) ,
\label{eq:transmitted_signal}
\end{equation}
where $P_{u,n}$ is the transmitted power for user $u$. 
It is worth noting that the BS reckons a user to be either a communication user (CU) or a sensing user (SU). As shown in Fig.~\ref{fig:scenario}, a CU is switched to an SU when its AoD estimation suffers from high uncertainty. Therefore, $\dot{s}_{u,n}(t)$ denotes either the data stream targeted for the CU, or is a basic radar waveform that contains no useful information, and $\mathrm{E}\{|\dot{s}_{u,n}(t)|^2\} = 1$. We assume that $\dot{s}$ for different users are uncorrelated, i.e., $\mathrm{E}\{\dot{s}_{u,n}(t)\dot{s}^*_{q,n}(t)\} = 0$, for $q\neq u$. 

It is to be noted that this work assumes block fading, and the index of TTI, $n$, is sometimes omitted for simplicity.

\subsection{Communication Model}
Within the $n$-th TTI, the signal received by the $u$-th user at time instant $t$ is
\begin{equation}
\small
\begin{aligned}
    \tilde{y}_{u,n}(t) = \boldsymbol{h}_{u,n} ^H \boldsymbol{s}_n(t) + \omega ,
\end{aligned}
\end{equation}
where $\omega \sim \mathcal{CN}(0,\sigma_{\omega}^2)$ is the zero-mean thermal noise with noise power $\sigma_{\omega}^2$ received at the user. The channel response between the BS and the $u$-th UE  
is expressed as
\begin{equation}
\small
    \boldsymbol{h}_{u,n} = \sum_{l=1}^{L} a_{l}(d_{u,n}) \boldsymbol{a}(\varphi_{u,n,l})  ,
\end{equation}
where $L$ denotes the number of propagation paths, $\varphi_{u,n,l}$ denotes the AoD of the $l$-th path and $a_{l}$ includes the amplitude and the phase of the path and it is a function of $d_u$, i.e., the distance from BS to the user. Specifically, $a_{1}(d_{u,n}) = \sqrt{N_t} v_c e^{j2 \pi d_{u,n}} /(4\pi d_{u,n} f_c)$ denotes the coefficient of the dominant line-of-sight (LoS) component. $\boldsymbol{a}(\theta) = \frac{1}{\sqrt{N_t}} [1, e^{j \theta}, \cdots, e^{j (N_t-1)\theta}]^T$ is the transmit steering vector, where the normalized AoD $\theta = \pi \sin \Tilde{\theta}$ is normalized from physical AoD $\Tilde{\theta}$. 

The average SINR for the $u$-th UE at TTI $n$ is given by
\begin{equation}
\small
    \gamma_{u,n} = \frac{P_{u.n}|\boldsymbol{h}_{u,n} ^H \boldsymbol{f}_{u,n}|^2}{\sum_{\substack{i=1, i \neq u}}^{U} |\sqrt{P_{i,n}} \boldsymbol{h}_{u,n}^H \boldsymbol{f}_{i,n} |^2 + \sigma_{\omega}^2} .
\end{equation}

\subsection{Sensing Model}


The transmitted signal impinges on the users and is reflected back to the sensing receiver equipped at the BS.
The reflected echoes can be written as~\cite{liu_integrated_2022}
\begin{equation}
\small
\begin{aligned}
   & \boldsymbol{y}_n(t) 
    =  \sum_{u=1}^{U} \beta_{u,n} e^{j2\pi \mu_{u,n}t}\boldsymbol{a}(\varphi_{u,n}) \boldsymbol{a}^H(\varphi_{u,n}) \boldsymbol{s}_n(t) + \boldsymbol{w}_n(t) ,
\end{aligned}
\label{eq:sensing_received_signal}
\end{equation}
where $\varphi_{u,n} \triangleq \varphi_{u,n,1}$ denotes the AoD of LoS path, $\beta_{u,n} = N_t \sqrt{v_c^2 \sigma_{rcs}/f_c^2 / (4\pi)^3/d_{u,n}^4}, u=1, \cdots, U$ denotes the reflection coefficient of the reflectors that includes two-way path loss and radar cross section (RCS) $\sigma_{rcs}$, $\mu_{u,n}$ denotes the Doppler frequency of the $u$-th user at TTI $n$,
and $\boldsymbol{\omega}_n(t)$ is assumed to be independent, zero-mean complex Gaussian with known covariance matrix $\mathbf{R}_{\omega} =\sigma_{\omega}^2 \mathbf{I}$. Considering the mono-static configuration and $N_r = N_t$, we assume the sensing echoes have equal AoD and angle of arrival. 

In such a co-located deployment of BS transmitter and sensing receiver, self-interference from the leakage of transmitted signal undermines reflected echoes. Moreover, the presence of fixed clutter in the environment is additional interference that hinders extracting sensing information. These interference sources can be mitigated by estimating their channels by leveraging the full information on the transmitted signal and perfectly known users' positions during a pilot transmission \cite{nayak_drl_2024} or considering hardware isolation \cite{zhuo_2024_multibeam_isac}. Fixed clutters can be solved by either multiple target localization techniques or by adaptive methods such as Capon’s beamformer \cite{bekkerman_target_2006}. Hence, in this work, we assume the self-interference and clutter interference from the environment are suppressed to the level of the background noise floor.

\subsection{Sensing - AoD Estimation}
The BS receives a composite echo reflected from multiple users, among which the BS identifies each user and estimates their AoDs in order to do beam management. Although the signal directions can be obtained by sophisticated methods, such as MUltiple SIgnal Classification (MUSIC), it is difficult for the BS to associate the classified directions with users without relying on user uplink feedback information. 
Nevertheless, the co-located deployment allows the sensing receiver to use the duplicates of the transmitted signal associated with a specific user to extract its information \cite{chen_multiuser_2023}.

Let us consider the $q$-th user. Adopting a matched filter with a delayed and Doppler-shifted version of $\dot{s}_{q,n}(t)$, one can estimate the Doppler frequency and delay as
\begin{equation}
 \small
    \left\{\hat{\mu}_{q,n},\hat{\tau}_{q,n} \right\}= \underset{\mu, \tau}{\arg \max} \left| \int_0^{\Delta T} \boldsymbol{y}_{n}(t) \dot{s}_{q,n}^* (t-\tau) e^{-j2\pi \mu t} dt \right|^2 .
\label{eq:delay-doppler}
\end{equation}
The distance between the $q$-th user and the BS is obtained as $\hat{d}_{q} = \hat{\tau}_{q,n}v_c/2$. Compensating the received signal $\boldsymbol{y}_n(t)$ by the counterpart of $\dot{s}_{q,n}(t)$ with the estimated delay and Doppler, we have the processed signal of the $n$-th TTI
\begin{subequations}
\small
\begin{align}
   \boldsymbol{\eta}_{q,n} =& \frac{1}{\Delta T} \int_0^{\Delta T} \boldsymbol{y}_n(t) \dot{s}^*_{q,n}(t - \hat{\tau}_{q,n}) e^{-j2\pi \hat{\mu}_{q,n} t} dt   \\
    =&  \sqrt{P_{q,n}} \beta_{q,n} \left(\boldsymbol{a}^H (\varphi_{q,n}) \boldsymbol{f}_{q,n}\right)\boldsymbol{a}(\varphi_q)   
    \label{eq:separated_echo} \nonumber \\ 
     & +  \sqrt{P_{q,n}} \sum_{\substack{u=1 \\ u\neq q}}^{U}   \beta_{u,n}  (\boldsymbol{a}^H (\varphi_{u,n}) \boldsymbol{f}_{q,n} ) \boldsymbol{a}(\varphi_u) + \tilde{\boldsymbol{\omega}}_n(t) 
,
\end{align}
\end{subequations}
where 
the second term in \eqref{eq:separated_echo} is the clutter for estimating the desired $\varphi_q$,  and the noise 
$\tilde{\boldsymbol{\omega}}_n(t)$ follows $\mathcal{CN}(\boldsymbol{0}, \sigma_{\omega}^2/\Delta T \mathbf{I})$. Note that this clutter, which indicates multi-user interference, is different from fixed clutter in the environment. The cross-correlation of the baseband waveform of different users is neglected as they are uncorrelated for a large $\Delta T$.  

The clutter in the processed signal impacts the estimation performance. This work considers a large antenna array at the BS and also assumes that users are separated enough. In this case, $\alpha_{uq,n}$ approximates to 0. Thus, the clutter has a limited effect on the estimation performance. By estimating the AoD of each user individually, the AoD and user association problem can be solved.

The AoD can be estimated as
\begin{equation}
\small
    \hat{\varphi}_{q,n} = \arg \underset{\theta}{\max} ~~ \boldsymbol{a}^H(\theta) \boldsymbol{\eta}_{q,n} \boldsymbol{\eta}^H_{q,n} \boldsymbol{a}(\theta) .
\label{eq:aod_estimate}
\end{equation}
Specifically, the CRLB of the maximum likelihood estimate of $\varphi_{q,n}$ when $\alpha_{uq,n}=0$ can be obtained as \cite{liu_crb_2022}
\begin{equation}
\small
    \sigma_{{\varphi}_{q,n}}^2 \ge \frac{ \sigma_{\omega}^2 / (P_{q,n}| \boldsymbol{a}^H (\varphi_{q,n}) \boldsymbol{f}_{q,n}|^2)}{2 \Delta T\left(\left\|\dot{\dot{\boldsymbol{a}}}(\varphi)\right\|^2 + \boldsymbol{a}^H(\varphi)\dot{\dot{\boldsymbol{a}}}(\varphi) +  \dot{\dot{\boldsymbol{a}}}^H(\varphi)\boldsymbol{a}(\varphi) \right)} ,
\label{eq:crlb}
\end{equation}
where $\dot{\dot{\boldsymbol{a}}}(\varphi) = \frac{\partial  \boldsymbol{a}^2(\varphi)}{\partial \varphi^2}$ is the second derivative.

As indicated by \eqref{eq:crlb}, the performance of estimating the AoD of user $q$ is affected by the gain achieved in the transmit mode, i.e., $|\boldsymbol{a}^H (\varphi_q) \boldsymbol{f}_{q,n}|^2$. This gain decreases due to the beam coverage when the target user is not in the center of the beam. This aligns with the result obtained in \cite{MugenPeng2023}. Meanwhile, the estimation precision of the AoD implies the quality of the allocated communication beam.


\section{Beam Management Problem}

This work focuses on dynamic beam management for resolving beam drift in sensing-assisted communication systems. It updates beams without user feedback but only with the obtained sensing information.
The result presented above shows that the CRLB increases when the beam is not directed to the target. Hence, reusing communication beams for sensing \cite{deep_rl_liu_2024_wcnc, Chen2024} is prone to induce beam failure when the communication beam drifts. A wide sensing FoV helps relieve the problem. To do so, the authors in \cite{zhang_multibeam_2019} proposed a multi-beam ISAC scheme that generates one time-invariant communication beam and one fast-varying sensing beam sweeping the area of interest during a channel coherence time. However, this requires the transmitter to switch the beamforming coefficients quickly, decreasing energy efficiency. Instead, we consider using multiple beams simultaneously for sensing to broaden the FoV.

In this section, we first describe the objective function of the sensing-assisted communication system and then propose an AoD-based method for managing multi-beams.    

\subsection{Objective Function}

We consider $U$ data buffers at the BS. Let $[B_{1,0}, \cdots, B_{U,0}]$ be the vector of the number of data packets generated at the beginning of a frame for each user, where $B_{u,0} \sim \mathbb{B}(N, \mathrm{Pr}_u)$ and $\mathbb{B}$ denotes the Binomial distribution. The total number of packets at the beginning of the frame is defined as $B_{tot} = \sum_{u=1}^{U} B_{u,0}$. A frame lasts for $N$ TTIs. Accordingly, the number of packets remaining in the buffer of user $u$ at TTI $n$ is denoted as $B_{u,n}$. Specifically, for a TTI, we consider that the packets are successfully transmitted when the communication rate is larger than a threshold $c$, that is, 
\begin{equation}
    \small \log (1 + \gamma_{u,n}) \geq c. 
    \label{eq:communication_criterion}
\end{equation}

The ultimate goal is to maximize the probability of successful transmission or equivalently minimize the remaining packets in the buffer at the end of the frame. The objective function is written as
\begin{equation}
\small
    \text{minimize} \sum_{u=1}^{U} B_{u,N} .
\label{eq:final_goal}
\end{equation}
The resources that the BS needs to dynamically manage during each TTI include beams and power for each user. To achieve the objective in \eqref{eq:final_goal}, the BS is supposed to update the beams when they are drifting. 

\subsection{AoD-Based Method}
We propose a scheme in which the BS groups users to CU for communication and SU for updating their AoD estimates dynamically at each TTI. The challenge is how the BS determines which mode each user is in. Ideally, the type of each user should be decided according to their estimation error. However, this requires the knowledge of true AoDs. Since the precision of AoD estimation implies the quality of the communication beam, we use this information for resource management without needing specific AoD estimation.

\subsubsection{User and Beam allocation} Let $\lambda_u$ denote the user type of the $u$-th user, i.e., $1$ represents a CU, 0 represents an SU. It is decided according to the estimation precision, i.e., 
\begin{equation}
\small \lambda_u = \boldsymbol{1}_{(\sigma_{{\varphi}_{u}} \le th)}  ,
\label{eq:sensing_criterion}
\end{equation}
where $\boldsymbol{1}$ denotes the indicator function, $th$ is a predefined threshold, and $\sigma_{{\varphi}_{u,n}}$ is the standard deviation of the estimation error calculated from the CRLB.
\begin{figure}
    \centering
    \includegraphics[width=.38\textwidth]{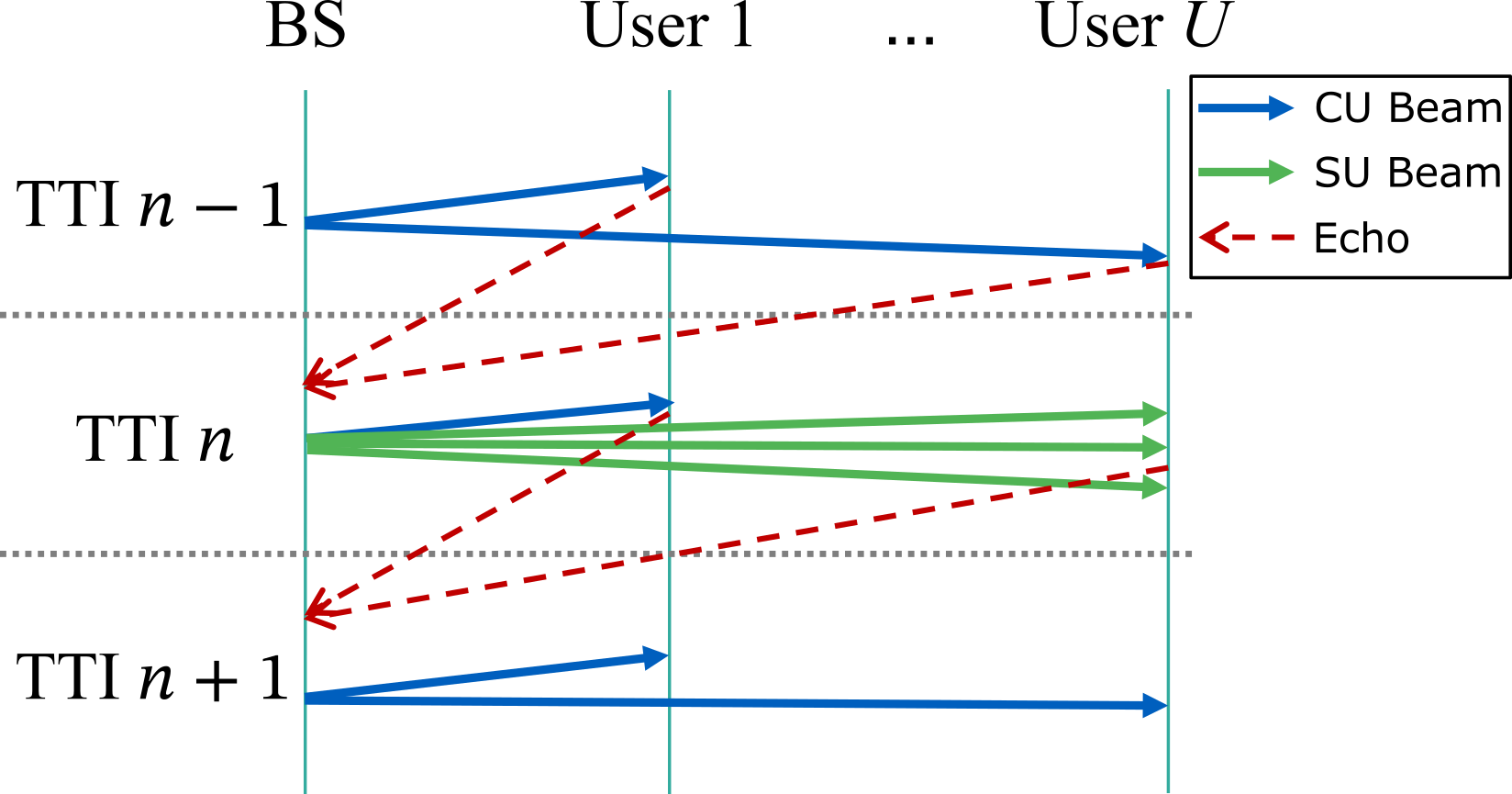}
    \caption{Sensing-Assisted Communication Protocol}
    \label{fig:protocol}
\end{figure}

The beam for a CU is set according to AoD estimate, that is, $\boldsymbol{f}_{u,n} = [\tilde{\mathbf{F}}]_{:,\hat{\varphi}'_{u,n-1}}$. The prime sign denotes that the estimated angles are approximated to the grid such that the beam is selected from the DFT codebook.
Accordingly, the beam for an SU is 
$\boldsymbol{f}_{u,n} =  \sum_{\hat{\varphi}'_{u,n-1} \in \{ \hat{\varphi}_{u,n-1} \pm \sigma_{{\varphi}_{u,n-1} \}} }[\tilde{\mathbf{F}}]_{:,\hat{\varphi}'_{u,n-1}}$.

It is noted that calculating CRLB requires genie AoD. The AoD-based method uses the approximated CRLB by substituting AoD estimates. Meanwhile, determining user type based on AoD estimation error and allocating beams according to CRLB is referred to as \emph{AoD-Genie} method. We consider it as an upper bound, which is evaluated in Section \ref{sec:simulation}.

The dynamic beam allocation protocol is exemplified in Fig. \ref{fig:protocol}. At the $n$-th TTI, the user beams are selected based on the echoes received during the $(n-1)$-th TTI. The $U$-th user is apart from the center of beam $\boldsymbol{f}_{U,n}$. Then, at TTI $n$, the BS allocates multiple beams for user $U$ to obtain a better estimation precision.


\subsubsection{Power allocation}\label{sec:power_allocation}
After the user type and beams during TTI $n$ are determined, the BS calculates the power allocation. 
The requirement of communication in \eqref{eq:communication_criterion} can be simplified to $\xi_c = \frac{|\boldsymbol{h}_u ^H \boldsymbol{f}_{u,n} |^2}{\sigma_{\omega}^2} \ge \frac{2^c-1}{P} \triangleq \iota_c$, 
and that of sensing in \eqref{eq:sensing_criterion} is simplified to $\xi_s = \frac{ \beta^2_{u,n} \big|\boldsymbol{a}^H (\varphi_q) \boldsymbol{f}_{u,n}\big|^2 \Delta T }{\sigma_{\omega}^2} \ge \frac{th^2}{P}\triangleq \iota_s$, where $\iota_s, \iota_c$ are the threshold for satisfying the requirements of sensing and communication respectively\footnote{The power for SU is equally distributed to the multiple beams. Optimizing the power to different beams will be studied in future work}.
Within a TTI, the objective function concerning power allocation is given by
\begin{equation}
\small
    \arg \max_{P_u} \sum _{u=1}^U \boldsymbol{1}_{\lambda_u P_u \frac{\xi_c}{\iota_{c}} + (1-\lambda_u)P_u \frac{\xi_s}{\iota_{s}} \ge 1} .
\end{equation}
This recasts the problem as maximizing the number of users whose signal strength is larger than a threshold, which is solved by the water-filling algorithm \cite{he_water-filling_2013}.
Note that $\xi_s$ and $\xi_c$ are approximated using the estimates of AoD in  \eqref{eq:aod_estimate} and of distance in  \eqref{eq:delay-doppler}.

\section{DRL-assisted Beam Management}
DRL is a technique where an agent employs deep neural networks to map observations to optimal actions. Proximal
policy optimization (PPO) \cite{ppo_2017} is a popular DRL algorithm that
prevents the policy from having a large update and performs
comparable or better than other DRL algorithms. 
In this paper, we use the PPO algorithm \cite{ppo_2017} to train the agent, i.e., BS. The BS observes the environment and allocates resources for all users. However, the action space is hybrid, containing discrete actions for user type and beam, and continuous actions for power level. We propose to simplify the agent by only taking discrete actions. Once the actions are set, the power allocation is calculated as in Section \ref{sec:power_allocation}.

Details on observation, actions, and reward functions are:

\emph{Observation:} 
The estimation performance is a function of the beamforming gain $|\boldsymbol{a}^H(\varphi_q) \boldsymbol{f}_{q,n}|^2$, as shown in  \eqref{eq:crlb}. This beamforming gain can be obtained at the BS by calculating the beamforming output 
\begin{equation}
\small
\boldsymbol{b}_{u,n} = \mathrm{E}\{|\boldsymbol{f}_{u,n}^H \boldsymbol{Y}_n|^2 \} 
\approx \beta_u^2 P_{u,n} |\boldsymbol{f}_{u,n}^H \boldsymbol{a}({\varphi_u})|^4  +
\sigma_n^2 ,
\end{equation}
where $\boldsymbol{Y}_n = [\boldsymbol{y}_{n}(1), \cdots, \boldsymbol{y}_n(\Delta T)]$ is the received echo within TTI $n$ and the expectation is taken over time. 
This is equivalent to measuring the power in the transmitted direction, which implies the beam allocation quality.
 
The state of the DRL model is the number of packets remaining in the buffer, the total number of packets for the frame, and the beamforming output with respect to the transmitted beams of each user. The state is written as
\begin{equation}
\begin{aligned}
    s_n  = \{&[B_{1}, \cdots, B_{U}],  B_{tot}, [\boldsymbol{b}_{1,n-2}, \cdots, \boldsymbol{b}_{U,n-2}], \\&[\boldsymbol{b}_{1,n-1}, \cdots, \boldsymbol{b}_{U,n-1}], [\boldsymbol{b}_{1,n}, \cdots, \boldsymbol{b}_{U,n}]\} \in \mathcal{S}.
\end{aligned}
\end{equation}
The BS restores the beamforming output of the past 3 TTIs. The historical beamforming output helps the BS learn the user's moving direction.

\emph{Action:}
The agent action is to determine user type and beams, which is given by
\begin{equation}
   a_n =   \{[\lambda_1, \cdots, \lambda_U], [\mathcal{C}_{1}, \cdots, \mathcal{C}_{U}]\} \in \mathcal{A} .
\end{equation}
where $\lambda_u$ is user type of the $u$-th user, and $\mathcal{C}_{u}$ denotes the beam indices that are allocated for user $u$.
The beamforming weight is computed as $\boldsymbol{f}_u = \sum_{i\in\mathcal{C}_u} \tilde{\mathbf{F}}_{i,:}$.

It is difficult to determine the action space dimension for $\mathcal{A}$ because each user has one or multiple beams depending on its type. One option is to assign every beam to one of the $U$ users or stay idle. This yields the action space dimension of $N_t^{U+1}$. In the mmWave spectrum, the number of beams $N_t$ is high leading that the convergence of the learning algorithm is hardly guaranteed, and large memory is required for computation. We shrink the action space as follows.
For the user $u$, there are two cases depending on the user type in the $(n-1)$-th TTI.

Case 1: $\lambda_{u,n-1} = 1$, and beam index is $\mathcal{C}_{u, n-1}$
\begin{itemize}
    \item $n$ TTI, $\lambda_{u,n} = 1$, the beam $\mathcal{C}_{u, n} = \mathcal{C}_{u, n-1}$
   \item $n$ TTI, $\lambda_{u,n} = 0$, the beam can be 
   \begin{itemize}
      \item $\mathcal{C}_{u, n} = \mathcal{C}_{u, n-1} + [-2,-1,0]$
     \item $\mathcal{C}_{u, n} = \mathcal{C}_{u, n-1} + [0, 1, 2]$
    \end{itemize}
\end{itemize}

Case 2: $\lambda_{u,n-1} = 0$, and beam index is $\mathcal{C}_{u, n-1}$ 
\begin{itemize}
   \item $n$ TTI, $\lambda_{u,n} = 1$, the beam $\mathcal{C}_{u, n}$ is obtained from the estimation result
   \item $n$ TTI, $\lambda_{u,n} = 0$, the beam $\mathcal{C}_{u, n}$ needs to cover more area
    \begin{itemize}
      \item $\mathcal{C}_{u, n} = \min(\mathcal{C}_{u, n-1}) + [-2,-1,0]$
      \item $\mathcal{C}_{u, n} = \max(\mathcal{C}_{u, n-1}) + [0, 1, 2]$
    \end{itemize}
\end{itemize}
When a user is a sensing user, the BS allocates some adjacent beams of its previous TTIs to enlarge its sweeping area. 
As a result, the action space dimension decreases to $3^U$.

\emph{Reward:}
The reward of each step is given by
\begin{equation}
\small
r^c = \left(1 + \exp \Big(\frac{B_{tot} - \sum_{u=1}^{U}B_u }{B_{tot}}\Big) \right) \sum_{u=1}^{U} \lambda_u \boldsymbol{1}_{\log(1+\gamma_{u,n}) >c} ,
\end{equation} 
where the first multiplicative term is set to encourage the agent to transmit more packets.
Even though the sensing criterion is not included in the reward function, the sensing performance is implicitly exhibited in the long term reward because it will affect the subsequent communications. 

The DRL-assisted method is initialized with that the agent knows the best beams towards each user and receives the echoes as the initial observation. In each step, the agent observes the state to select the action based on the policy. The policy network consists of 2 fully connected layers with 64 units which are also shared with the critic network. PPO algorithm trains the policy by using a clip function in the surrogate objective function to prevent the policy from having a large variance \cite{ppo_2017}. The clipping parameter is set to 0.2. 




\section{Simulation Results}\label{sec:simulation}

\begin{table}
    \centering
    \caption{SIMULATION PARAMETERS}
    \resizebox{\columnwidth}{!}{%
    \begin{tabular}{c c c}
    \hline \hline
    Description &Symbol &Value \\ 
    \hline
    BS antennas &$N_t$& 32 \\
    Frequency &$f_c$& 28 GHz \\
    Tx Power &$P_t$& 15 dBm \\
    Noise Power& $\sigma_{\omega}^2$& $-109$ dBm\\
    Rician Factor& $K$ &10 dB\\
    Time duration of a TTI& $dt$& 10 ms\\
    Beamwidth - Sensing Threshold & $\theta_{bw}$ & $\approx 4^{\circ} $ \\
    Rate - Communication Threshold & $c$ & 4 bits/s/Hz \\
   Radar Cross Section  & $\sigma_{rcs} $ & 25 $m^2$ \cite{Motomura_rcs_2018}\\
    Scenario Dimensions & -- & $[100 \times 100]$ \\
    Number of UEs &$U$& 2 \\
    Initialized Packet Number & $B_{u,0}$ & $\sim$ $\mathbb{B}(100,0.6)$\\
    User Speed &$v$&$\sim \mathcal{N}(\bar{v},4), \bar{v}=15,20,25,30$ \\
    Number of TTI per Frame &$N$ & 100 \\
    Number of training steps & -- & $10^5$ \\
    Number of testing episodes & -- & 2000 \\
    \hline \hline
    \end{tabular}%
   }
    \label{tab:exp_para}
\end{table}
\begin{figure}[t!]
    \vspace{-3mm}
    \centering
    \includegraphics[width=0.39\textwidth]{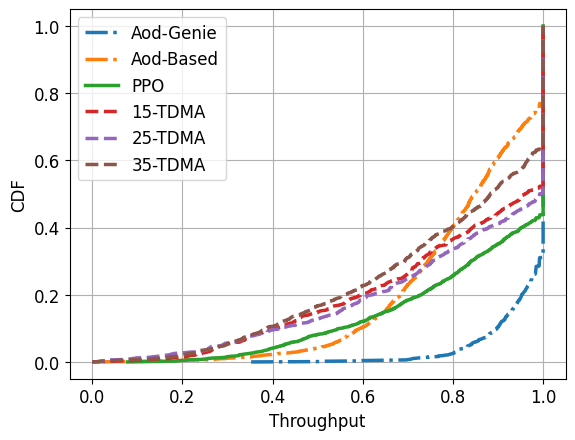}
    \caption{CDF of normalized throughput when users averaged speed is 20 m/s}
    \label{fig:throughput_cdf}
\end{figure}

This section presents the simulation result of the proposed multi-beam management in a sensing-aided vehicular communication system. The simulation parameters are summarized in Table. \ref{tab:exp_para}. Specifically, the considered scenario is a $100 \text{m} \times 100 \text{m}$ area with a BS situated in the center, as illustrated in Fig.~\ref{fig:street}. The light grey area is streets with four crosses. LoS can be guaranteed between the BS and the road such that the wireless channel is considered Rician fading with the Rician K factor being 10 dB. A vehicle coming to a cross can turn to any of the other three directions with equal probability. Vehicles' speed is set to follow the normal distribution.
The 3dB beamwidth for a linear array with $N_t$ antennas is approximately $\theta_{bw} \approx 2 \arcsin\{\frac{1.2}{N_t}\}$ \cite{luo_multibeam_2020}, which is used as the threshold for deciding user type in the AoD-based method. The rate threshold for the communication criterion is set to be 4 bits/s/Hz.
With respect to training the PPO-based method, $10^5$ steps are used, and a step is equivalent to a TTI. The duration of a TTI is 10 ms. There are 12 sets of vehicle initial positions. During testing, vehicle positions are initialized randomly on the road. 

The normalized throughput of every frame is used as the metric to evaluate different methods. It is given by
$\text{Thp} \triangleq \frac{B_{tot} - \sum_{u=1}^{U} B_{u,N} }{B_{tot}}$. At the end of the frame, the remaining packets are discarded.
\emph{$X$-TDMA ($X$ Time Division Multiple Access)} is selected as the benchmark. This scheme imitates how the standardized protocol operates. It is independent of the agent's environment. The BS assigns a sensing slot to all users before every $X$ communication slot. 
\emph{AoD-Genie} is the upper bound. This method requires genie information on AoDs. It uses the estimation error to determine user types and CRLB to determine beams. 
All the considered methods use the water-filling power allocation presented in Section \ref{sec:power_allocation}. 

\begin{figure}[t!]
    \centering
    \includegraphics[width=0.39\textwidth]{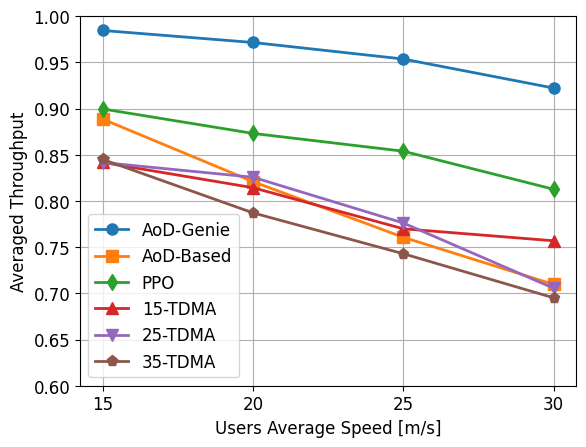}
    \caption{Comparison of averaged throughput as a function of user speed among different methods}
    \label{fig:throughput_speed}
\end{figure}

Fig.~\ref{fig:throughput_cdf} shows the comparison of the cumulative distribution function (CDF) of normalized throughput when users move at an average speed of 20 m/s. The AoD-Genie is the upper bound. The performance of the AoD-based method has a performance degradation because it has an information loss on the AoDs, and it only uses statistics of estimation error for deciding beams. The trained PPO achieves a better throughput compared to conventional TDMA fashion and AoD-based method in high percentile.   
The performance of $X$-TDMA is highly related to system kinematics. In Fig. \ref{fig:throughput_cdf}, 25-TDMA has the best performance compared with the other two because 15-TDMA has too-frequent sensing, which occupies communication resources, while 35-TDMA does not update beams timely. This phenomenon can also be observed in Fig. \ref{fig:throughput_speed}, which shows the variation of averaged throughput as a function of user speed. The averaged throughput decreases as users move faster because beam drifting appears more often. Overall, the PPO-based approach outperforms the conventional methods. Therefore, it is robust against different moving speeds.

\section{Conclusion}
The dynamic beam management for alleviating beam drifting in multi-user sensing-aided communication is investigated. Since the reflected echoes received at the BS sensing receiver imply the quality of the beam allocation, the BS utilizes only these reflected echoes for beam management. A DRL-assisted approach that maximizes system throughput is presented. For comparison, a heuristic AoD-based method using estimation precision is also proposed. The clear performance gain on throughput shows that the DRL-assisted approach alleviates beam drifting in time compared to conventional TDMA fashion and the AoD-based method and is robust to different user speeds. Future work will study the beamforming architecture that better manages inter-beam interference. In addition, the power allocation of multiple beams will be optimized to further enhance the performance.

\vspace{-3mm}





\bibliographystyle{IEEEtran}
\bibliography{centric}

\begin{thebibliography}{10}
\providecommand{\url}[1]{#1}
\csname url@samestyle\endcsname
\providecommand{\newblock}{\relax}
\providecommand{\bibinfo}[2]{#2}
\providecommand{\BIBentrySTDinterwordspacing}{\spaceskip=0pt\relax}
\providecommand{\BIBentryALTinterwordstretchfactor}{4}
\providecommand{\BIBentryALTinterwordspacing}{\spaceskip=\fontdimen2\font plus
\BIBentryALTinterwordstretchfactor\fontdimen3\font minus \fontdimen4\font\relax}
\providecommand{\BIBforeignlanguage}[2]{{%
\expandafter\ifx\csname l@#1\endcsname\relax
\typeout{** WARNING: IEEEtran.bst: No hyphenation pattern has been}%
\typeout{** loaded for the language `#1'. Using the pattern for}%
\typeout{** the default language instead.}%
\else
\language=\csname l@#1\endcsname
\fi
#2}}
\providecommand{\BIBdecl}{\relax}
\BIBdecl

\bibitem{liu_integrated_2022}
F.~Liu, Y.~Cui, C.~Masouros, J.~Xu, T.~X. Han, Y.~C. Eldar, and S.~Buzzi, ``Integrated {Sensing} and {Communications}: {Toward} {Dual}-{Functional} {Wireless} {Networks} for {6G} and {Beyond},'' \emph{IEEE J. Sel. Areas Commun}, vol.~40, no.~6, pp. 1728--1767, Jun. 2022.

\bibitem{xue_survey_2024}
Q.~Xue, C.~Ji, S.~Ma, J.~Guo, Y.~Xu, Q.~Chen, and W.~Zhang, ``A {Survey} of {Beam} {Management} for {mmWave} and {THz} {Communications} {Towards} {6G},'' \emph{IEEE Commun. Surv. Tutor}, vol.~26, no.~3, p. 1520–1559, Feb. 2024.

\bibitem{zhuo_2024_multibeam_isac}
Y.~Zhuo, T.~Mao, H.~Li, C.~Sun, Z.~Wang, Z.~Han, and S.~Chen, ``Multi-beam integrated sensing and communication: State-of-the-art, challenges and opportunities,'' \emph{IEEE Commun. Mag.}, vol.~62, no.~9, pp. 90--96, 2024.

\bibitem{Zhang2021BeamDrift}
J.~Zhang and C.~Masouros, ``Beam drift in millimeter wave links: Beamwidth tradeoffs and learning based optimization,'' \emph{IEEE Trans. Commun.}, vol.~69, no.~10, pp. 6661--6674, 2021.

\bibitem{MugenPeng2023}
J.~Zhang, S.~Yan, and M.~Peng, ``Joint beam alignment and resource allocation for multi-user mmwave integrated sensing and communication systems,'' \emph{IEEE Trans. Veh. Technol.}, pp. 1--16, 2023.

\bibitem{chen_multiuser_2023}
K.~Chen, C.~Qi, and O.~A. Dobre, ``\BIBforeignlanguage{en}{Multiuser {Beam} {Tracking} and {Target} {Detection} in {Integrated} {Sensing} and {Communication}},'' in \emph{\BIBforeignlanguage{en}{IEEE ICC2023}}.\hskip 1em plus 0.5em minus 0.4em\relax Rome, Italy: IEEE, May 2023, pp. 5743--5748.

\bibitem{liu_radar-assisted_2020}
F.~Liu, W.~Yuan, C.~Masouros, and J.~Yuan, ``Radar-{Assisted} {Predictive} {Beamforming} for {Vehicular} {Links}: {Communication} {Served} by {Sensing},'' \emph{IEEE Trans. Wireless Commun.}, vol.~19, no.~11, pp. 7704--7719, Nov. 2020.

\bibitem{Chen2024}
L.~Chen, K.~Liu, Z.~Zhang, and B.~Li, ``{Beam} {Selection} and {Power} {Allocation}: {Using} {Deep} {Learning} for {Sensing}-{Assisted} {Communication},'' \emph{IEEE Wireless Commun. Lett.}, vol.~13, no.~2, pp. 323--327, 2024.

\bibitem{deep_rl_liu_2024_wcnc}
Y.~Liu, S.~Zhang, X.~Li, Y.~Huang, Y.~Fang, and H.~Cao, ``{Deep} {Reinforcement} {Learning}-{based} {Beamforming} {Design} in {ISAC}-{assisted} {Vehicular} {Networks},'' in \emph{2024 IEEE WCNC}, 2024.

\bibitem{Li_drl_bm_2022}
L.~Yan, X.~Fang, S.~Li, Y.~Li, and Q.~Xue, ``{DRL} {Based} {Beam} {Management} for {Joint} {Sensing} and {Communications} in {HSR} {mmWave} {Wireless} {Networks},'' in \emph{2022 IEEE 95th VTC2022-Spring}, 2022.

\bibitem{nayak_drl_2024}
N.~Nayak, S.~Kalyani, and H.~A. Suraweera, ``A {DRL} {Approach} for {RIS}-{Assisted} {Full}-{Duplex} {UL} and {DL} {Transmission}: {Beamforming}, {Phase} {Shift} and {Power} {Optimization},'' \emph{IEEE Trans. Wireless Commun.}, vol.~23, no.~10, pp. 14\,652--14\,666, 2024.

\bibitem{bekkerman_target_2006}
I.~Bekkerman and J.~Tabrikian, ``Target {Detection} and {Localization} {Using} {MIMO} {Radars} and {Sonars},'' \emph{IEEE Trans. Signal Process.}, vol.~54, no.~10, pp. 3873--3883, Oct. 2006.

\bibitem{liu_crb_2022}
F.~Liu, Y.-F. Liu, A.~Li, C.~Masouros, and Y.~C. Eldar, ``Cramér-rao bound optimization for joint radar-communication beamforming,'' \emph{IEEE Trans. Signal Process.}, vol.~70, pp. 240--253, 2022.

\bibitem{zhang_multibeam_2019}
J.~A. Zhang, X.~Huang, Y.~J. Guo, J.~Yuan, and R.~W. Heath, ``Multibeam for {Joint} {Communication} and {Radar} {Sensing} {Using} {Steerable} {Analog} {Antenna} {Arrays},'' \emph{IEEE Trans. Veh. Technol.}, vol.~68, no.~1, pp. 671--685, Jan. 2019.

\bibitem{he_water-filling_2013}
P.~He, L.~Zhao, S.~Zhou, and Z.~Niu, ``Water-{Filling}: {A} {Geometric} {Approach} and its {Application} to {Solve} {Generalized} {Radio} {Resource} {Allocation} {Problems},'' \emph{IEEE Trans. Wireless Commun.}, vol.~12, no.~7, pp. 3637--3647, July 2013.

\bibitem{ppo_2017}
J.~Schulman, F.~Wolski, P.~Dhariwal, A.~Radford, and O.~Klimov, ``{Proximal} {Policy} {Optimization} {Algorithms},'' \emph{CoRR}, vol. abs/1707.06347, 2017.

\bibitem{Motomura_rcs_2018}
T.~Motomura, K.~Uchiyama, and A.~Kajiwara, ``Measurement results of vehicular rcs characteristics for 79ghz millimeter band,'' in \emph{2018 IEEE WiSNet}, 2018, pp. 103--106.

\bibitem{luo_multibeam_2020}
Y.~Luo, J.~A. Zhang, X.~Huang, W.~Ni, and J.~Pan, ``Multibeam {Optimization} for {Joint} {Communication} and {Radio} {Sensing} {Using} {Analog} {Antenna} {Arrays},'' \emph{IEEE Trans. Veh. Technol.}, vol.~69, no.~10, pp. 11\,000--11\,013, Oct. 2020.

\end{thebibliography}

\end{document}